\documentstyle[preprint,aps]{revtex}
\begin{document}
\draft
\title{Comment on ``Anomalously Large Gap Anisotropy \\
in the a-b Plane of Bi$_{{\bf 2}}$Sr$_{{\bf 2}}
$CaCu$_{{\bf 2}}$O$_{{\bf 8+}{\delta}}$"}
\author{Kazumasa Miyake and Osamu Narikiyo}
\address{Department of Material Physics, Faculty of Engineering Science \\
Osaka University, Toyonaka 560, Japan}
\date{\today}
\maketitle

\narrowtext

In a recent Letter \cite{shen}, Shen and collaborators reported that
angle-resolved photoemission spectra (ARPES) in Bi-2212 cuprate superconductor
far below $T_{{\rm c}}$ has anomalously large anisotropy and then concluded
that the symmetry of Cooper pair is likely to be $d_{x^{2}-y^{2}}$, i.e.,
$\Delta_{{\bf k}}=\Delta(\cos k_{x}a-\cos k_{y}a)$ \cite{takahashi}.  In this
Comment we point out that the interpretation done by Shen {\it et al\/.},
where the shift of edge near Fermi level has been identified with the gap
$\Delta_{{\bf k}}$, is not unambiguous, and that a possibility of $d_{xy}$
pairing, i.e., $\Delta_{{\bf k}}=2\Delta\sin k_{x}a\sin k_{y}a$, cannot be
ruled out.

The ARPES intensity is proportional to $A({\bf k},\omega)\equiv-1/\pi\cdot
{\rm Im}G^{\rm R}({\bf k},\omega)$ given in the superconducting state as
follows:
\widetext
\begin{equation}
A({\bf k},\omega)=z_{{\bf k}}{-\omega-{\tilde {\xi}}_{{\bf k}}
\over2{\tilde E}_{{\bf k}}}\cdot{1\over\pi}\Biggl\{{{\tilde{\gamma}}_{{\bf k}}
\over\bigl(\omega+{\tilde E}_{{\bf k}}\bigr)^{2}
+{\tilde{\gamma}}_{{\bf k}}^{2}}-{{\tilde{\gamma}}_{{\bf k}}
\over\bigl(\omega-{\tilde E}_{{\bf k}}\bigr)^{2}
+{\tilde{\gamma}}_{{\bf k}}^{2}}\Biggr\}+A_{\rm inc}({\bf k},\omega),
\label{intensity}
\end{equation}
\narrowtext
where $z_{{\bf k}}$ is the renormalization amplitude, ${\tilde E}_{{\bf k}}
\equiv(\Delta_{{\bf k}}^{2}+{\tilde {\xi}}_{{\bf k}}^{2})^{1/2}$ and
${\tilde {\gamma}}_{{\bf k}}$ being the dispersion and damping of
quasiparticles.  The first term of Eq.\ (\ref{intensity}) gives coherent
contribution due to quasiparticles and the second term,
$A_{\rm inc}({\bf k},\omega)$, represents incoherent background.
In the photoemission experiment ($\omega<0$), the first term in the coherent
part, if it exists, gives dominant contribution.

First of all, it is remarked that the peak position of the coherent component
is shifted downward by amount of $(\Delta_{{\bf k}}^{2}
+{\tilde {\xi}}_{\bf k}^{2})^{1/2}+{\tilde {\xi}}_{\bf k}$ when the
superconductivity sets in.  The {\it maximum} gap $2\Delta$ at $T=0$ is
related to the transition temperature $T_{{\rm c}}$ as
$2\Delta(0)\simeq3.5k_{{\rm B}}T_{{\rm c}}$ if it is estimated by the weak
coupling treatment assuming $d_{x^{2}-y^{2}}$ pairing.  Since
$T_{{\rm c}}=78$K and then the {\it maximum} gap $2\Delta(0)
\simeq2.7\times10^{2}$K, such a shift of peak position should be detectable
there far below $T_{{\rm c}}$ within their relative experimental resolution
of $\sim5$meV \cite{shen}.

Secondly, the damping rate of quasiparticles in the normal state is given
roughly as ${\tilde{\gamma}}\simeq [(k_{{\rm B}}T)^{2}
+(\omega/\pi)^{2}]^{1/2}$ \cite{ito}.  Therefore, narrowing of the coherent
peak around $\omega\sim-$30meV, which are the case in a typical example of
Fig.\ 1 in Ref.\ \cite{shen}, seems to be detectable when the temperature is
decreased from 85K down to 20K provided wave number {\bf k} is located at the
position such that $\Delta_{{\bf k}}=0$.

\newpage
Thirdly, the spectral weight just below the Fermi level in the superconducting
state is half of that in the normal state if $|{\tilde {\xi}}_{{\bf k}}|$ is
much less than $\Delta_{{\bf k}}$.  This is because half of the spectral
weight of normal quasiparticles is transferred to above the Fermi level due
to gap formation.  Namely, ARPES intensity of coherent part in the
superconducting state near the Fermi level is rather harder to observe than
in the normal state if the gap $\Delta_{{\bf k}}$ takes maximum there.

Now we encounter difficulty in interpretation of Ref. \cite{shen}.  The ARPES
at the location $A$ on $\Gamma$-${\bar M}$ line in Fig.\ 1 does not show
visible shift of peak position at all, but exhibits narrowing of the peak
width suggesting a separation of coherent peak and incoherent background.
If the gap had maximum at $A$, the peak should have shifted downward about
15meV in contrary to the observation in Fig.\ 1.  Therefore, as an alternative
interpretation, it seems more appropriate to consider that the gap
$\Delta_{{\bf k}}$ vanishes along the $k_{x}$-axis ($k_{y}=0$), parallel to
Cu-O bond, and that we observed the narrowing of the coherent peak at
$\omega={\tilde {\xi}}_{{\bf k}}$  which had been located just below the Fermi
level at ${\tilde {\xi}}_{\bf k}\sim-$30meV.  The present interpretation is
also consistent with the so-called aging effect reported in Ref.\ \cite{shen},
because the inhomogeneity of the surface caused by deficit of oxygen or so is
expected to give an excess of damping of quasiparticles leading to broadening
of the coherent peak.

On the other hand, at the location $B$ on $\Gamma$-$Y$ line, where the
coherent component is hardly seen in the normal state (note that the maximum
intensity at $A$ and $B$ in the normal state are almost the same while the
incoherent background at $B$ is larger than that at $A$ by 30-40\%), there
appears a peak around $\omega\sim-$30meV far below $T_{{\rm c}}$ without
showing apparent shift of the edge near the Fermi level.  If the gap vanished
at $B$ as interpretted in Ref.\ \cite{shen}, the narrowing of the coherent
peak should have been observed.  The intensity profile at $B$ in Fig.\ 1
rather looks as if new peak is added to the incoherent background when the
temperature is decreased far below $T_{{\rm c}}$.  So it seems more natural to
consider that B is located just at or above the Fermi level and a new peak
appeares around $\omega=-{\tilde E}_{{\bf k}}\simeq-$30meV but with a reduced
weight $z_{{\bf k}}(-{\tilde {\xi}}_{\bf k}+{\tilde E}_{{\bf k}})/
2{\tilde E}_{{\bf k}}<z_{{\bf k}}/2$.

\newpage
In conclusion, ARPES data of Ref.\ \cite{shen} are also understood even if we
assume $d_{xy}$ pairing which has gap vanishing along $\Gamma$-${\bar M}$
line and reaching maximum around $B$.  Theoretical arguments favoring such
pairing in cuprates have been put forth in various contexts
\cite{littlewood,narikiyo}.  Kane {\it et al.}\/ has also suggested in a very
recent Letter \cite{kane} that results of scanning tunneling microscope in
Bi-2212 sample are understood on the basis of $d_{xy}$ pairing \cite{sato}.

\end{document}